\documentclass[a4paper]{jpconf}
\usepackage{graphicx}
\begin{document}
\title{The Top Mass: \\Interpretation and Theoretical Uncertainties}

\author{Andr\'e H.~Hoang}

\address{University of Vienna, Faculty of Physics, Boltzmanngasse 5, A-1090 Wien, Austria\\[2mm]
	Erwin Schr\"odinger International Institute for Mathematical Physics,
	University of Vienna, Boltzmanngasse 9, A-1090 Vienna, Austria}

\ead{andre.hoang@univie.ac.at \hfill UWTHPH-2014-33}

\begin{abstract}
Currently the most precise LHC measurements of the top quark mass are determinations of the top quark mass parameter of Monte-Carlo (MC) event generators reaching uncertainties of well below $1$~GeV. However, there is an additional theoretical problem when using the MC top mass $m_t^{\rm MC}$ as an input for theoretical predictions, because a rigorous relation of $m_t^{\rm MC}$ to a renormalized field theory mass is, at the very strict level, absent. In this talk I show how - nevertheless - some concrete statements on $m_t^{\rm MC}$ can be deduced assuming that the MC generator behaves like a rigorous first principles QCD calculator for the observables that are used for the analyses. I give simple conceptual arguments showing that in this context $m_t^{\rm MC}$ can be interpreted like the mass of a heavy-light top meson, and that there is a conversion relation to field theory top quark masses that requires a non-perturbative input. The situation is in analogy to B physics where a similar relation exists between experimental B meson masses and field theory bottom masses.
The relation gives a prescription how to use $m_t^{\rm MC}$ as an input for theoretical predictions in perturbative QCD. 
The outcome is that at this time an additional uncertainty of about $1$~GeV has to be accounted for. I discuss limitations of the arguments I give and possible ways to test them, or even to improve the current situation.
\end{abstract}

\section{Introduction}

After the discovery of the Higgs boson, making new more precise measurements of the properties of the Standard Model particles and their interactions is the major aim of current and upcoming LHC analyses, next to their search for new particles. In this context the measurements of the top quark mass play an important role because its large size influences many quantitative and conceptual aspects within the Standard Model as well as in models for physics beyond the Standard Model. Currently, the most precise measurements come from direct reconstruction of the top quark decay final states and use templates (or kinematic fits) of data and MC output to obtain the top quark mass value that describes the data best. The first combination of data from LHC and Tevatron using the available analysis channels recently obtained the result
$m_t^{\rm MC} = 173.34 \pm 0.76$~\cite{ATLAS:2014wva}.  
Extrapolations for this method indicate that an uncertainty of below $0.5$~GeV might be achieved during the upcoming LHC $14$~TeV runs once an integrated luminosity of $300\,{\rm fb}^{-1}$ has been accumulated~\cite{cernreportmass}. For an alternative prospective discussion I also refer to~\cite{Juste:2013dsa}. 

Unfortunately, the result cannot be used directly as an input for precise NLO/NNLO theoretical predictions because the measured quantity is the top mass parameter of the MC event generators which is not a renormalized field theory mass. It has been frequently assumed that $m_t^{\rm MC}$ should be basically equal to the renormalized top quark pole mass $m_t^{\rm pole}$. The corresponding arguments were, however, at best vague and the corresponding uncertainty has never been quantified systematically. In this talk I discuss a number of simple thoughts on the physics of $m_t^{\rm MC}$ 
starting from the assumption that the MC generators can, to a good approximation, carry out first principles QCD computations - at least for the set of observables used for the top mass measurements.\footnote{This assumption is needed at the starting point because otherwise almost nothing can be said about $m_t^{\rm MC}$.}
This starting point appears justified due to the good description MC generators give for these and many other observables.
I hope that these thoughts clarify the issue and might make statements on uncertainty estimates more transparent. 

The outcome is that $m_t^{\rm MC}$ can be related to specific low-scale short-distance mass schemes with an uncertainty of around $1$~GeV. These mass schemes can then be used in theoretical predictions or converted to other mass schemes. It is in principle possible to improve the situation by additional theoretical analyses. Such theoretical analyses can also provide non-trivial tests of the whole argumentation. More could be said to flesh out the story from a more global point of view, which would also illuminate the connection to other top mass determinations based on more inclusive measurements such as the total cross section or recent studies of top-antitop plus jet final states as well as the relation of the top mass to thresholds or kinematic endpoints.\footnote{The methods based on the total and on the top-antitop plus jet cross section allow for the measurement of a renormalized field theory top quark mass. These methods are less sensitive to the top quark mass than direct reconstruction and currently yield uncertainties slightly above $2$\,GeV. I refer to~\cite{Juste:2013dsa,Moch:2014tta,Adomeit:2014yna} for details and a discussion of alternative top quark mass determination methods.}
But due to lack of time I will focus here exclusively on the MC mass problem.

\section{Comments on the top quark mass in QCD}
\label{sec:masses}

\begin{figure}[h]
	\hspace{10mm}
	\includegraphics[width=8pc]{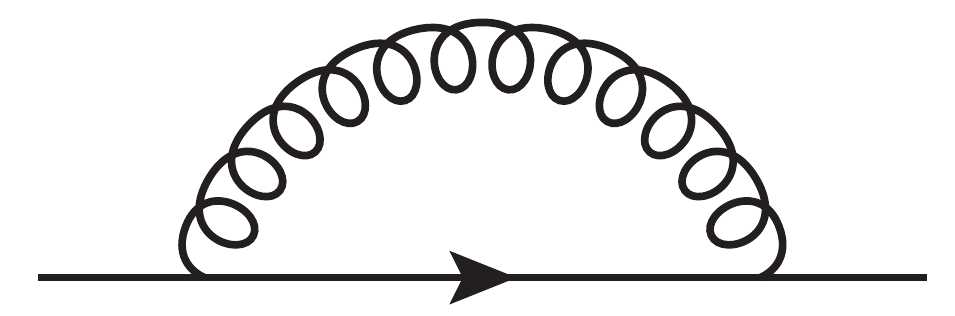}\hspace{2pc}%
	\begin{minipage}[b]{17pc}\caption{\label{fig:selfenergy}Top quark self energy at NLO. }
	\end{minipage}
\end{figure}

It is impossible to talk about the matter without discussing some field theoretic and (un)physical aspects of the top quark mass. The top quark mass is a renormalized quantity of the QCD Lagrangian due to UV (large momentum) divergences that arise in the top self energy Feynman diagrams (see Fig.~\ref{fig:selfenergy}) and need to be absorbed into the mass. Additional finite contributions of the self energy can be absorbed as well in this procedure, and different choices for the finite contributions define different top mass schemes. 

One very well known scheme is the {\bf $\overline{\mbox{MS}}$ scheme}, where only pure UV divergences are absorbed into the top mass. This scheme is mostly used for processes involving high energies of order of the mass or above and the corresponding mass $\overline m(\mu)$ is renormalization scale dependent like the strong coupling $\alpha_s(\mu)$. Physically, the $\overline{\mbox{MS}}$ scheme is (conceptually and numerically) very far away from the notion of a physical particle mass relevant for the top quark kinematics. The parameter $\overline m(\mu)$ should be thought more as being a coupling for a heavy quark-antiquark correlation and is therefore a very good scheme for parameterizing the top Yukawa coupling.   

The other very well known scheme is the top quark {\bf pole scheme}, where all UV and finite contributions of the self energy are absorbed into the mass in the on-shell limit $q^2=(m_t^{\rm pole})^2$. The pole scheme would correspond to our intuitive notion of the top quark's physical rest mass, but this notion and its connection to the top quark's kinematic properties are physically limited because of confinement. The important conceptual issue in this context is that the pole mass scheme is based on the perception of the self energy diagram being a meaningful physical quantity. This is, however, only so for momenta above $1$~GeV (which is the energy value I use in this talk for what theorists call "hadronization scale") because perturbation theory breaks down for energies below $1$~GeV. This concerns the contributions from the top quark self energy as well as the contributions from the other Feynman diagrams relevant for the process considered (describing gluon exchange between quarks and gluon radiation).
For the top quark pole mass this is a problem because the pole mass renormalization prescription prevents a specific set of contributions from momenta below $1$\,GeV to cancel among all Feynman diagrams. This problem is known as the pole mass {\rm renormalon} problem~\cite{Beneke:1998ui} and leads to a deterioration of the perturbative series in the pole scheme, which subsequently makes a determination $m_t^{\rm pole}$ with a precision of better than $1$\,GeV intrinsically impossible. 
The way how this problem becomes manifest in practice depends strongly on the typical scale of the observable under consideration with respect to the scale $1$\,GeV. For typical scales much larger than $1$~GeV (e.g.\ for total or very inclusive cross sections) the problem might not become obvious even at NNLO, while for typical scales closer to $1$~GeV (e.g.\ threshold cross sections, bound states or resonance problems) the issue might so severe that the use of the pole mass scheme is out of question (as I show in Sec.~\ref{sec:mesons}). Let me remark that the large decay width of the top quark is not affecting anything of what I just said, as was shown in~\cite{Smith:1996xz}. This is because the renormalon problem is not physical problem, but one that is "homemade" by an inappropriate theoretical use of perturbation theory.  

We see that thinking about a top mass determination with an uncertainty smaller than $1$~GeV based on templates and kinematic fits, requires devising a more suitable quark mass scheme. Such a scheme should absorb corrections from the heavy quark self energy as well, but only those coming from scales above $1$~GeV. Indeed, a number of such schemes, called "low-scale short-distance masses", have been devised in the context of top and B physics (see e.g.~\cite{Hoang:2000yr,Fleming:2007xt,Hoang:1998ng,Gambino:2013rza,Pineda:2001zq}). All these schemes differ slightly in the way they are implemented, but they generically depend on the scale $R$ (which is sometimes called {\it subtraction scale}). Only self-energy contributions above the scale $R$ are absorbed into these mass definition, while self-energy contributions below $R$ are left unabsorbed such that they can cancel with contributions from other diagrams. I would like to mention in particular the top quark {\bf MSR mass} $m^{\rm MSR}(R)$ devised in \cite{Hoang:2008yj} which is constructed such that it formally interpolates in a smooth way between the pole and the $\overline{\mbox{MS}}$ mass,
\begin{eqnarray}
m^{\rm MSR}(R) & \stackrel{R\to 0}{\longrightarrow} & m^{\rm pole}\,,\\\nonumber
m^{\rm MSR}(R=\overline m(\overline m)) & = & \overline m(\overline m)\,,
\end{eqnarray}
and for with the R-dependence is described by a renormalization group evolution equation. In Ref.~\cite{Hoang:2008yj} it was pointed out that all short-distance masses associated to a given $R$ scale have somewhat comparable values, so we can view $m^{\rm MSR}(R)$ in the following set of arguments as a representative for all low-scale short-distance masses with the subtraction scale $R$. For simplicity I will therefore only talk about the MSR mass for the subsequent discussion.
For a choice of $R\sim 1$~GeV the MSR mass is a scheme which is as close as possible to the pole mass (i.e. carrying all the kinematic information for reconstruction {\it that can be described by perturbation theory above $1$~GeV}) but at the same time avoids the renormalon problem. For the sake of being specific I will in the following mostly talk about the MSR mass for $R=1$~GeV, but I stress that all scales in the range from $1$ to around $3$~GeV might be reasonable choices.

\begin{figure}[h]
	\hspace{16mm}
	\includegraphics[width=13pc]{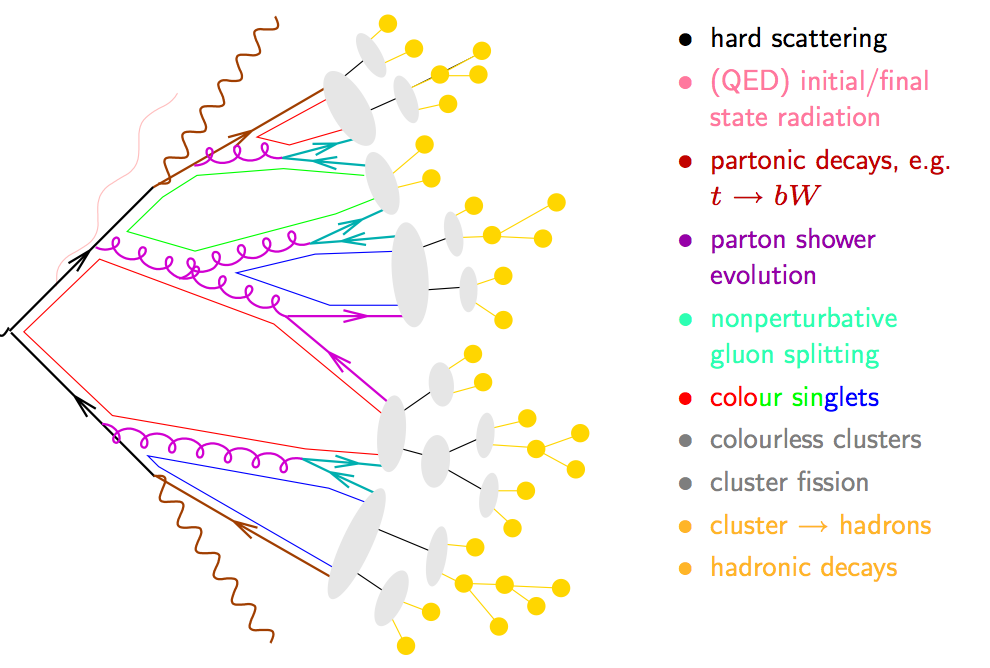}\hspace{2pc}%
	\begin{minipage}[b]{14pc}\caption{\label{fig:MC} Illustration of MC components for the final state interactions in top-antitop production. Figure from D. Zeppenfeld.}
	\end{minipage}
\end{figure}

\section{The top quark mass in the MC event generator}

Let us now turn our attention to the role of the top quark mass in a MC event generators, see Fig.~\ref{fig:MC}. MC event generators contain matrix elements computed in perturbation theory describing the hard interactions. For top quark production the matrix elements describe the initial state parton annihilation and the initial production of top quarks and potential additional hard partons. The MC top quark mass parameter is the mass in the top propagator prior to the top quark decay. Attached to the hard final state partons of the matrix elements is the parton shower evolution which describes the top decays and the continued splitting into higher multiplicity partonic states having subsequently lower virtualities. The splitting probabilities are calculated from perturbative QCD. Some MC's use other shower evolution parameters such $p_T$ or energy-weighted angles, but the principle is the same.

\begin{figure}[h]
	\hspace{20mm}
	\includegraphics[width=12pc]{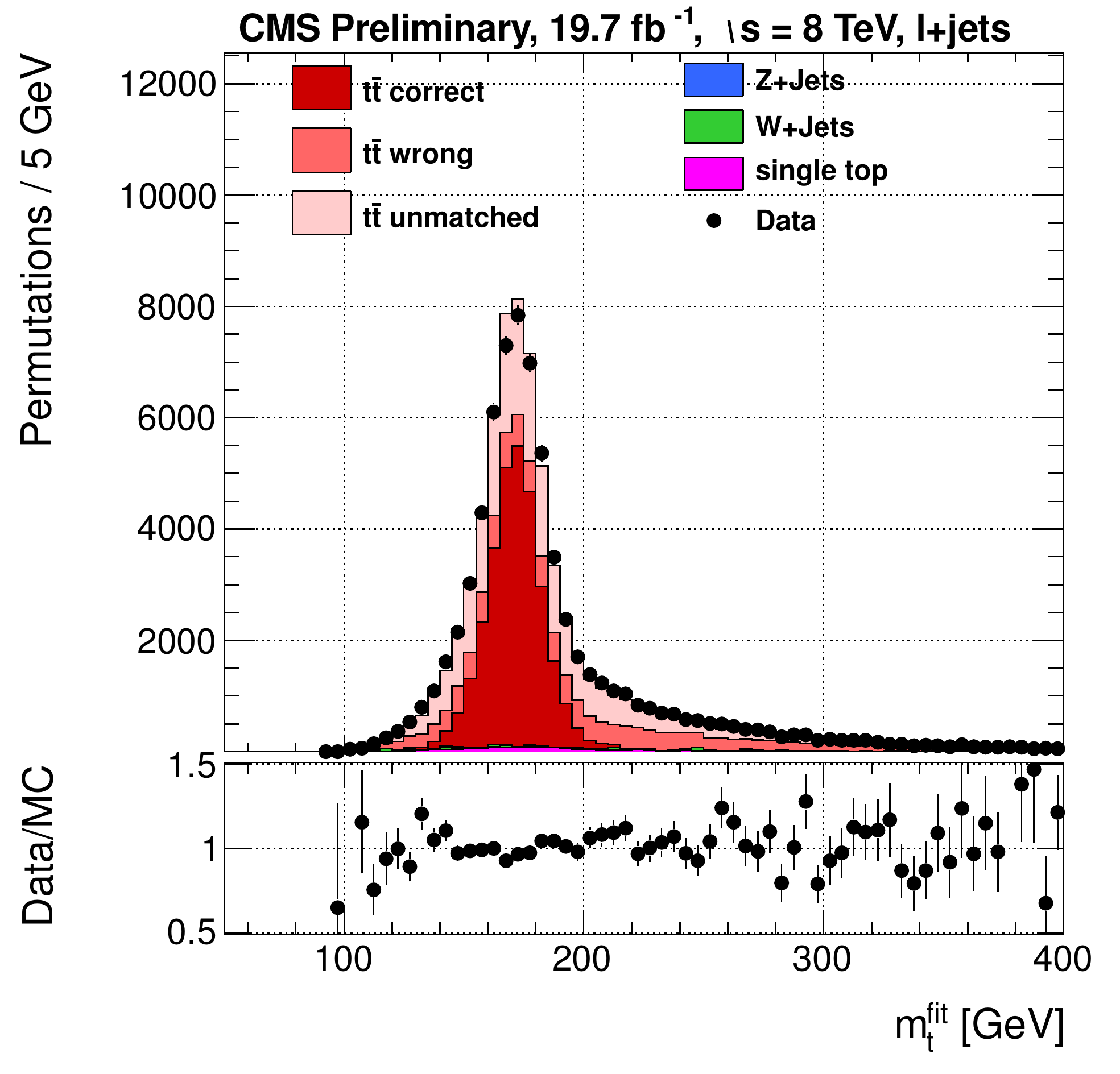}\hspace{2pc}%
	\begin{minipage}[b]{14pc}\caption{\label{fig:cms}Top invariant mass distribution from kinematic fits obtained in the CMS analysis~\cite{CMS:2014ima}. }
	\end{minipage}
\end{figure}

For the matter of my argumentation let me view the parton shower evolution as a way to sum up perturbative corrections (valid to leading order for the top decay and valid to an approximate leading logarithmic accuracy for soft-collinear splitting) coming from energies between the hard scale contained in the matrix element and the virtuality scale. The property of the parton shower that is important for the conclusions below is that it does not account for any top quark self-energy corrections. At some point in the evolution the parton shower reaches particle virtualities of $1$~GeV where the perturbative description of the splitting process begins to fail. At this point, which is called the shower cut $\Lambda_s$, the parton shower terminates and a hadronization model takes over to turn the partons into observable hadrons or jets of hadrons that can be observed. 
Conceptually, the MC event generator follows the logic of factorization which states that to a good approximation the various energy (or virtuality) ordered processes can be described separately. For all MC event generators, however, the implementations (or tunes) of the hadronization models are strongly tied to details of the implementations of the parton-shower which are described by a probabilistic Markov chain. 
It is the result of the tunig procedure (to a set of well known reference processes) that the resulting quality of MC generators to describe hadron level data, particularly in the soft-collinear limit is frequently so high, and that one does not have to worry about let's say missing NLO virtual corrections in the parton shower. The question on how to interpret the MC top mass $m_t^{\rm MC}$ from the first principles QCD point of view is therefore also a {\it question on whether MC generators are more like very good models or more related to first principles QCD}. 

From this set up it is straightforward to see how the templates used for fitting the MC top mass depend on the different components of the MC event generator. Let's take the reconstructed top invariant mass distribution as a concrete example, see Fig.~\ref{fig:cms}.
The hard matrix elements describe the production rate of the hard configuration of the top quark events. So to a good approximation the hard matrix elements affect only the overall norm of the top invariant mass distribution, but not the details of its form or where it is peaked. The reconstructed top invariant mass distribution calculated by the MC generator is therefore determined by the value of the MC top mass, the parton shower and the hadronization model. The MC top mass determines the overall location of the mass range where the distribution is peaked. The parton shower and the hadronization model provide additional effects that further modify the shape and the peak location, and the interplay of both
is essential for MC top mass measurements with uncertainties below $1$ GeV.\footnote{The experimental strategies for the definition of the templates often include ways to minimize the contributions from soft radiation. This does, however, not remove the importance of parton shower and hadronization model for the intrinsic meaning of the MC top mass which is observable independent.} In this context the parton shower is responsible for perturbative QCD corrections and the hadronization model is responsible for non-perturbative QCD corrections. 

Finally we are at the point where we should address the question of the physical interpretation of the MC top mass parameter in the templates. Recalling that the perturbative corrections summed up by the parton shower evolution do not account for any top quark self-energy contribution we must consider them being absorbed into the MC mass parameter, but only for energy scales above the shower cut $\Lambda_s=1$~GeV. For scales below $1$~GeV the partonic degrees of freedom are not used any more and the non-perturbative hadronization model is employed. The infrared issues known from perturbative QCD from scales below $1$~GeV never arise in this context and there are also no perturbative contributions to the mass parameter coming from this region. This already tells us that $m_t^{\rm MC}$ is not the top quark pole mass. Recalling our discussion in Sec.~\ref{sec:masses} we see that the MC top mass $m_t^{\rm MC}$ has, as far as its perturbative contributions from the self-energy are concerned, the same features as the low-scale short-distance masses such as the MSR mass $m^{\rm MSR}(R)$ with $R$ equal to the shower cut of $1$~GeV. The difference between the MC top mass $m_t^{\rm MC}$ and the MSR mass $m_t^{\rm MSR}(R=\Lambda_s)$ is mainly related to the contributions coming from scales below $\Lambda_s=1$~GeV: 
The MSR mass is employed within full QCD calculations (using dimensional regularization) where perturbation theory contributes corrections from below the scale $R=1$~GeV and hadronization effects are included based on factorization. The MC mass, on the other hand, is employed exclusively within a hadronization model.

From this we can conclude that the difference between the two masses (which both are quantities with dimension of energy) is a {\it non-perturbative contribution} of around $1$~GeV:
\begin{eqnarray}
\label{eq:MCMSR}
	& & m_t^{\rm MC} \, = \, m_t^{\rm MSR}(R=1\,\mbox{GeV})+\Delta_{t,\rm MC}(R=1\,\mbox{GeV})\,,\\
	\nonumber
	& &\Delta_{t,\rm MC}(1\,\mbox{GeV}) \, \simeq \, {\cal O}(1\,\mbox{GeV})\,.
\end{eqnarray}
The term $\Delta_{t,\rm MC}(R)$ is actually a sum of a perturbative and a non-perturbative contribution. The perturbative component controls the scheme-dependence on the RHS of Eq.~(\ref{eq:MCMSR}), e.g.\ related to variations in the choice of $R$ around $1$~GeV. Since $m_t^{\rm MSR}(R)$ is unambiguously defined as a perturbative QCD series currently known up to ${\cal O}(\alpha_s^3)$~\cite{Hoang:2008yj} , the RHS of Eq.~(\ref{eq:MCMSR}) can be reliably converted to any other short-distance mass scheme. So it is possible to express the RHS in terms of $m_t^{\rm MSR}(R)$ for another $R$ value, with a corresponding scheme change in $\Delta_{t,\rm MC}(R)$, since $m_t^{\rm MC}$ is $R$-independent. -- But only for $R$ close to $1$~GeV the size of $\Delta_{t,\rm MC}(R)$ is around $1$~GeV.

Relation~(\ref{eq:MCMSR}) and the statement about the size of $\Delta_{t,\rm MC}(1\,\mbox{GeV})$ are the central results arising from our thoughts. They parameterize -- in I believe the best possible way -- the meaning of the MC top quark mass and what the numerical value from~\cite{ATLAS:2014wva} means within the context of QCD from first principles.
They have been derived before in Ref.~\cite{Hoang:2008xm} using a similar set of thoughts based on a calculation of the reconstructed invariant mass distribution for boosted top quarks in $e^+e^-$ annihilation~\cite{Fleming:2007xt,Fleming:2007qr}.\footnote{In Ref.~\cite{Hoang:2008xm} $\Delta_{t,\rm MC}(R\approx 1\,\mbox{GeV})$ was treated as an uncertainty, parameterized by a variation of $R$, and the relation corresponding to~(\ref{eq:MCMSR}) was quoted as $m_t^{\rm MC} \, = \, m_t^{\rm MSR}(3^{+6}_{-2}\,\mbox{GeV})$.} 
The conceptual reliability of these thoughts and thus of the relations~(\ref{eq:MCMSR}) themselves is mostly based on the interplay of the MC generator's parton shower, which is at best a leading-logarithmic approximation, and the MC generator's tuned hadronization model. We are back to the question whether the MC generatos are more like models or more first principles QCD. One might argue that there is no need to worry about this (too much), because the MC generators are known to describe the templates used for the reconstruction method much better than a theoretical leading-logarithmic approximation could ever do, and because of the observed consistency of the various MC mass determinations.
In any case, there certainly are conceptual and maybe even numerical limitations in relation~(\ref{eq:MCMSR}) that still need to be tested theoretically. I come back to this in Sec.~\ref{sec:final}, where I discuss how $\Delta_{t,{\rm MC}}(R=1\,\mbox{GeV})$ might be determined.

\section{Top mesons, connection to B physics and current estimates of $\Delta(R=1~\mbox{GeV})$}
\label{sec:mesons}

The conceptual relation encoded in~(\ref{eq:MCMSR}) has an interesting and well-known analogy. This is the difference between the mass of a T meson (i.e.\ bound states of a top quark and a light anti-quark) and $ m_t^{\rm MSR}(R=1\,\mbox{GeV})$. For this relation heavy quark theory~\cite{Manohar:2000dt} states that $\Delta$ is non-perturbative and has a size of around $1$~GeV just as given in~(\ref{eq:MCMSR}). The analogy it not to be understood as a physical statement since the MC top mass parameter is not parameterizing a meson mass (not even in the limit of stable top quarks where mesons could in principle form and also because the top is colored while mesons are not). Rather, the relation should be understood as a practical expression which carries all the conceptual aspects we have discussed above. I have to admit, however, that the idea of associating physical particle properties to top quark propagating in the MC simulations is not uncommon in the community, so that the analogy to the T meson should not be too uncomfortable.

Based on the idea that we might consider $m_t^{\rm MC}$ to be like a T meson mass we can now draw a connection to B physics, where the analogous relation for a B meson mass reads
\begin{eqnarray}
\label{eq:bottom}
m_B \, = \, m_b^{\rm MSR}(1\,\mbox{GeV})+\Delta_{b,B}(R=1\,\mbox{GeV})\,,\qquad \Delta_{b,B}(1\,\mbox{GeV})\simeq{\cal O}(1\,\mbox{GeV})\,,
\end{eqnarray}
and heavy quark theory~\cite{Manohar:2000dt} further states that $\Delta_{b,B}$ is quark mass independent. 
Since $\Delta_{b,B}$ is non-perturbative and quite hard to calculate from QCD, the sub-MeV precision in the experimentally known B meson masses cannot be transferred to a sub-MeV level precision for the bottom quark mass. In fact, $\Delta_{b,B}$ is so hard to calculate that current most accurate ways to determine the bottom mass are based on completely different methods reaching uncertainties at the level of $50$~MeV or somewhat below~\cite{Agashe:2014kda}. 
However, from the PDG we can determine $\Delta_{b,B}(1~\mbox{GeV})$ using the perturbative relation (known at ${\cal O}(\alpha_s^3)$) of the quoted bottom 1S mass $m_b^{\rm 1S}=4780\pm 66$~MeV to  
$m_b^{\rm MSR}(1\,\mbox{GeV})$. The perturbative conversion uncertainty is negligible and the results are shown in Tab.~\ref{tabmb} for different B mesons. We see that the value for $\Delta_{b,B}(1\,\mbox{GeV})$ range between around $0.49$ and $0.95$~GeV which is compatible with the generic estimate of $1$\,GeV. Because, due to heavy quark symmetry, these numbers would be unchanged for the corresponding T mesons we might take them as a guideline for values we can expect for $\Delta_{t,\rm MC}(1\,\mbox{GeV})$. For comparison I have also shown the corresponding numbers for $R=2$~GeV.

\begin{table}[h]
	\caption{\label{tabmb}Some B mesons masses, MSR masses $m_b^{\rm MSR}(1\,\mbox{GeV})$ and $m_b^{\rm MSR}(2\,\mbox{GeV})$ from $m_b^{\rm 1S}=4780\pm 66$\,MeV\,\cite{Agashe:2014kda}, and corresponding values for $\Delta_{b,B}$. All in units of MeV, $\alpha_s(m_Z)=0.1184$.}
	\begin{center}
		\begin{tabular}{cccccc}
			\br
			 $m_b^{\rm MSR}(1\,\mbox{GeV})$  & $m_b^{\rm MSR}(2\,\mbox{GeV})$  &$m(B^0)$ & $m(B^*)$ & $m(B_1^0)$ & $m(B_2^*)$\\
			\mr
			$4795\pm 69$ & $4571\pm 69$ & $5279.58 \pm 0.17$  & $5325.2 \pm 0.4$  & $5724 \pm 2$ & $5743 \pm 5$\\
			\mr\mr
			$\Delta_{b,B}(1\,\mbox{GeV})$ & & $485\pm 69$ & $530\pm 69$ & $929\pm 69$ &  $948\pm 69$\\
			\mr
			& $\Delta_{b,B}(2\,\mbox{GeV})$ & $709\pm 69$ & $754\pm 69$ & $1153\pm 69$ & $1172\pm 69$\\
			\mr
		\end{tabular}
	\end{center}
\end{table}

Interestingly for the B mesons masses we can also illustrate in a striking way the problems of using the pole mass scheme:
If one attempts to determine $\Delta_{b,B}$ for the pole mass scheme, the outcome is order-dependent and incoherent because of the pole mass renormalon problem. Here the renormalon problem in manifest in the perturbative relation between the 1S mass (which is a low-scale short-distance mass) and the pole mass that breaks down. Since a coherent numerical analysis with properly quoted uncertainties is impossible in this context, let me just quote that for $m_b^{\rm 1S}=4.78$\,GeV the pole mass values at (1,2,3) loops read ($5.01$, $5.10$, $5.74$)\,GeV, 
($4.88$, $4.99$, $5.18$)\,GeV and ($4.85$, $4.94$, $5.08$)\,GeV
converting for the renormalization scales $\mu=1$, $2$ and $3$\,GeV, respectively. The spread of values is by far larger than the uncertainty in the 1S mass disqualifying the pole mass in this context.

\section{Some practical consequences}

From the thoughts above a number of practical consequences emerge:
\begin{itemize}
	\item Using event generators with NLO vs.\ LO hard matrix elements for MC top mass measurements from templates (or kinematic fits) based on direct reconstruction does to a good approximation not affect the interpretation of $m_t^{\rm MC}$. This is because the more essential ingredient is the parton shower (plus hadronization model tuned to the same reference processes) that is used in a leading approximation in all current MC generators.
	\item The MC top mass $m_t^{\rm MC}$ in principle depends on details of the implementation of the parton shower and the tuned hadronization model in each MC generator. Current reconstruction analyses based on different MC's seem to indicate that the differences are not big. It should be mentioned, however, that sizeable differences might arise for MC generators with NLO improved parton showers, since they would also entail a significant change in the hadronization model tunes.
	\item Relation~(\ref{eq:MCMSR}) can also be formulated using other quark mass schemes on the RHS of the equality. The term $\Delta_{t,\rm MC}$ then depends on the scheme, and all schemes can be converted by perturbation theory. For low-scale mass schemes closely related to $m_t^{\rm MSR}(1\,\mbox{GeV})$ the size of $\Delta_t$ is expected to be minimal.
	\item Each MC event generator's MC top mass has a unique meaning and its measured value should not depend on the observable one considers. So it should also not matter whether measurements are based on data from hadron or lepton colliders.
	\item In experimental top quark mass analyses which aim to determine the $\overline{\rm MS}$ mass $\overline m_t(\overline m_t)$, e.g.\ from inclusive cross sections, MC event generators can be used to determine experimental efficiencies. In this context Eq.~(\ref{eq:MCMSR}) together with known perturbative relation between the $\overline{\rm MS}$ mass and $m_t^{\rm MSR}(1\,\mbox{GeV})$ should be employed to relate $\overline m_t(\overline m_t)$ and  $m_t^{\rm MC}$.	
	\item At this time we do not have more information on $\Delta_{t,\rm MC}(1\,\mbox{GeV})$ than shown in~(\ref{eq:MCMSR}). So at this time $\Delta_{t,\rm MC}(1\,\mbox{GeV})\sim 1$\,GeV should be taken as uncertainty. Because we are talking about a theoretical error a more precise statement is impossible. I stress that this uncertainty {\it is not} an additional uncertainty in $m_t^{\rm MC}$, but one in the relation of  $m_t^{\rm MC}$ to $m_t^{\rm MSR}(1\,\mbox{GeV})$. 
	\item One can also relate $m_t^{\rm MC}$ to the top quark pole mass using a relation similar to (\ref{eq:MCMSR}). However, making a proper estimate for the uncertainty associated to $\Delta_{t,\rm MC}$ is harder. This is because using the top quark pole mass in theory calculations can leads to a systematic bias that can vary case-by-case. In general, for the pole mass 
	$1$~GeV should be considered as a lower bound for the size of $\Delta_{t,\rm MC}$. The issue can be very severe in predictions for observables involving small energies (e.g.\ top pair threshold at a lepton collider) and might be milder for observables involving very high scales (e.g.\ total cross sections at large c.m.\ energies).     
\end{itemize}

\section{Outlook and possible improvement}
\label{sec:final}

The current situation could be improved by a determination of $\Delta_{t,\rm MC}(1\,\mbox{GeV})$. This can only be achieved by comparing {\it hadron level} predictions for templates such as the reconstructed top invariant mass distribution made by the MC event generator with corresponding {\it hadron level} first principle QCD calculations. So far first QCD calculations of such kind have been achieved already in the context of $e^+e^-$ collisions~\cite{Fleming:2007xt,Fleming:2007qr}. 
From the comparison one can calibrate the MC mass in terms of a well defined field theory mass. Such first principles QCD calculations must have control over the full top quark mass dependence and involve perturbative as well as non-perturbative components and are thus beyond the realm of purely perturbative QCD calculations regardless at which level they contain fixed-order corrections or logarithmic resummations. 
They might of course be also used directly to determine a well-defined field theory top quark mass from experimental data. However, calibrating the MC mass has a merit if the observable considered for the comparison is less suitable for data analysis, e.g.\ due to background or pile-up issues, but still highly sensitive to the MC mass. In the best of cases we might expect a precise determination of $\Delta_{t,\rm MC}(1\,\mbox{GeV})$ with a small error.

To conclude, I would like to reiterate that the conceptual basis of the MC top mass has intrinsic limitations that cannot be overcome - simply because MC event generators are not really first principles QCD computers. These limitations are related to the question how close MC event generators are to first principles QCD. These limitations will become manifest in the calibration analyses mentioned above e.g.\ by irreducible discrepancies to the first principles QCD predictions or by an observable-dependence of the MC mass calibrations, and they will set the ultimate precision to which $\Delta_{t,\rm MC}(1\,\mbox{GeV})$ can be determined. So the precision to which $\Delta_{t,\rm MC}(1\,\mbox{GeV})$ can be determined is a reflection of the conceptual meaning of the MC top quark mass. However, for the time until such theoretical analyses have been carried out thoroughly, $\Delta_{t,\rm MC}(1\,\mbox{GeV})$ has to be treated as an uncertainty.


\section*{References}
\begin{thebibliography}{9}
	
	\bibitem{ATLAS:2014wva} 
	[ATLAS and CDF and CMS and D0 Collaborations],
	arXiv:1403.4427 [hep-ex].
	
	\bibitem{cernreportmass}
	The CMS Collaboration, CMS physics analysis symmary, CMS PAS FTR-13-017
	
	\bibitem{Juste:2013dsa} 
	A.~Juste, S.~Mantry, A.~Mitov, A.~Penin, P.~Skands, E.~Varnes, M.~Vos and S.~Wimpenny,
	Eur.\ Phys.\ J.\ C {\bf 74}, no. 10, 3119 (2014)
	[arXiv:1310.0799 [hep-ph]].
	
	\bibitem{Moch:2014tta} 
	S.~Moch, S.~Weinzierl, S.~Alekhin, J.~Blumlein, L.~de la Cruz, S.~Dittmaier, M.~Dowling and J.~Erler {\it et al.},
	arXiv:1405.4781 [hep-ph].
	
	\bibitem{Adomeit:2014yna} 
	S.~Adomeit,
	arXiv:1411.7917 [hep-ex].
	
	\bibitem{Beneke:1998ui} 
	M.~Beneke,
	Phys.\ Rept.\  {\bf 317}, 1 (1999)
	[hep-ph/9807443].
	
	\bibitem{Smith:1996xz} 
	M.~C.~Smith and S.~S.~Willenbrock,
	Phys.\ Rev.\ Lett.\  {\bf 79}, 3825 (1997)
	[hep-ph/9612329].
	
	\bibitem{Hoang:2000yr} 
	A.~H.~Hoang, M.~Beneke, K.~Melnikov, T.~Nagano, A.~Ota, A.~A.~Penin, A.~A.~Pivovarov and A.~Signer {\it et al.},
	Eur.\ Phys.\ J.\ direct C {\bf 2}, 1 (2000)
	[hep-ph/0001286].
	
	\bibitem{Fleming:2007xt} 
	S.~Fleming, A.~H.~Hoang, S.~Mantry and I.~W.~Stewart,
	Phys.\ Rev.\ D {\bf 77}, 114003 (2008)
	[arXiv:0711.2079 [hep-ph]].
	
	\bibitem{Hoang:1998ng} 
	A.~H.~Hoang, Z.~Ligeti and A.~V.~Manohar,
	Phys.\ Rev.\ Lett.\  {\bf 82}, 277 (1999)
	[hep-ph/9809423].
	
	\bibitem{Gambino:2013rza} 
	P.~Gambino and C.~Schwanda,
	Phys.\ Rev.\ D {\bf 89}, 014022 (2014)
	[arXiv:1307.4551 [hep-ph]].
	
	\bibitem{Pineda:2001zq} 
	A.~Pineda,
	JHEP {\bf 0106}, 022 (2001)
	[hep-ph/0105008].
	
	\bibitem{Hoang:2008yj} 
	A.~H.~Hoang, A.~Jain, I.~Scimemi and I.~W.~Stewart,
	Phys.\ Rev.\ Lett.\  {\bf 101}, 151602 (2008)
	[arXiv:0803.4214 [hep-ph]].
	
	\bibitem{CMS:2014ima} 
	CMS Collaboration [CMS Collaboration],
	CMS-PAS-TOP-14-001.
	
	\bibitem{Hoang:2008xm} 
	A.~H.~Hoang and I.~W.~Stewart,
	Nucl.\ Phys.\ Proc.\ Suppl.\  {\bf 185}, 220 (2008)
	[arXiv:0808.0222 [hep-ph]].
	
	\bibitem{Fleming:2007qr} 
	S.~Fleming, A.~H.~Hoang, S.~Mantry and I.~W.~Stewart,
	Phys.\ Rev.\ D {\bf 77}, 074010 (2008)
	[hep-ph/0703207].
	
	\bibitem{Manohar:2000dt} 
	A.~V.~Manohar and M.~B.~Wise,
	Camb.\ Monogr.\ Part.\ Phys.\ Nucl.\ Phys.\ Cosmol.\  {\bf 10}, 1 (2000).
	
	\bibitem{Agashe:2014kda} 
	K.~A.~Olive {\it et al.}  [Particle Data Group Collaboration],
	Chin.\ Phys.\ C {\bf 38}, 090001 (2014)
	[arXiv:1412.1408].
	
\end{thebibliography}
\end{document}